# Magnetoelectric Interactions in Layered Composites of Piezoelectric Quartz and Magnetostrictive Alloys

G. Sreenivasulu,[1] V. M. Petrov,[1,2] L. Y. Fetisov,[1,3] Y. K. Fetisov,[1,4] and G. Srinivasan[1a]

[1] Physics Department, Oakland University, Rochester, MI 48307
[2] Institute for Information Systems, Novgorod State University, Veliky Novgorod, Russia
[3] Department of Physics, Moscow State University, Moscow 119991, Russia
[4] Moscow State Technical University of Radio Engineering, Electronics and Automation, Moscow 119454, Russia

**ABSTRACT**

Mechanical strain mediated magnetoelectric effects are studied in bilayers and trilayers of piezoelectric quartz and magnetostrictive permendur (P), an alloy of Fe-Co-V. It is shown that the magneto-electric voltage coefficient (MEVC), proportional to the ratio of the piezoelectric coupling coefficient to the permittivity, is higher in quart-based composites than for traditional ferroelectrics based ME composites. In bilayers of X-cut single crystal quartz and permendur, the MEVC varies from 1.5 V/cm Oe at 20 Hz to ~ 185 V/cm Oe at bending resonance or electromechanical resonance corresponding to longitudinal acoustic modes. In symmetric X-cut quartz-P trilayers, the MEVC ~ 4.8 V/cm Oe at 20 Hz and ~ 175 V/cm Oe at longitudinal acoustic resonance. Trilayers of Y-cut quartz and permendur show ME coupling under a shear strain with an MEVC that is an order of magnitude smaller than for longitudinal strain in samples with X-cut quartz. A model for low-frequency and resonance ME effects which allows for explicit expressions of MEVC and resonance frequencies is provided and calculated MEVC are in general agreement with measured values. Magneto-electric composites with quartz have the desired characteristics such as the absence of ferroelectric hysteresis and pyroelectric losses and could potentially replace ferroelectrics in composite-based magnetic sensors, transducers and high frequency devices.

[a] corresponding author: srinivas@oakland.edu

## 1. Introduction

The nature of strain mediated electromagnetic interactions in multiferroic composites of magnetostrictive and ferroelectric phases has attracted considerable attention in recent years [1-3]. Two types of magneto-electric (ME) coupling are of importance in such systems. (i) Direct ME effects: an ac magnetic field $\delta H$ applied to the composite produces a mechanical deformation in the ferromagnetic phase that in turn is coupled to the ferroelectric layer, resulting in an electric field $\delta E$ due to piezoelectric effect [1, 2]. The ME voltage coefficient $\alpha_E = \delta E/\delta H$ is a measure of the strength of direct ME interaction. (ii) Converse ME effects: An applied electric field $E$ giving rise to strain induced magnetization $M$ and is measured from data on $M$ vs $E$ or electric field tuning of ferromagnetic resonance [1-3]. Strong direct and converse ME interactions are reported for composites with ferrites, manganites, ferromagnetic metals or alloys for the magnetic phase and barium titanate, lead zirconate titanate (PZT), lead magnesium niobate-lead titanate (PMN-PT), lead zinc niobate-lead titanate (PZN-PT) or PVDF for the ferroelectric phase [1-5].

Here we report on the observation of strong direct ME effects in composites with piezoelectric quartz and a ferromagnetic alloy. Studies so far have focused primarily on composites with ferroelectric phases such as PZT or PMN-PT that in general show strong ME coupling due to high piezoelectric coupling coefficient $d$ [1-6]. But the ME coefficient $\alpha_E$ is directly proportional to the ratio $d/\varepsilon$ where $\varepsilon$ is the permittivity. Thus piezoelectrics with $d/\varepsilon$-values comparable to traditional ferroelectrics have attracted some attention very recently in this regard. Systems studied to-date includes AlN, ZnO, and lanthanum gallium tantalate (LGT) [6-9]. A giant low-frequency $\alpha_E$ and enhancement at bending and acoustic resonances are reported for composites of ferromagnetic alloys with AlN or single crystal LGT of composition $La_3Ga_{5.5}Ta_{0.5}O_{14}$ [7,9].

Here we provide details of our studies on direct-ME effects in bilayers and trilayers of single crystal quartz and permendur, a ferromagnetic alloy of Fe-Co-V. Since X-cut quartz has $d/\varepsilon$ ~ 0.04 m$^2$/VF (versus 0.017 and 0.035 m$^2$/VF for PZT and PMN-PT, respectively), one anticipates a relatively strong ME coupling in composites with quartz [10-13]. Data on ME response at low frequency, bending resonance and electromechanical resonance (EMR) are obtained for the composites. For quartz-permendur bilayers, $\alpha_E$ ~ 1.5 V/cm Oe at 20 Hz, 185 V/cm Oe at bending resonance and 190 V/cm Oe at longitudinal EMR. For symmetric trilayers of permendur-quartz-permendur, $\alpha_E$ ~ 4.8 V/cm Oe at 20 Hz and ~ 175 V/cm Oe at EMR. Similar ME measurements on composites with Y-cut quartz and permendur show near-zero $\alpha_E$ for in-plane strain and 0.45 V/cm Oe for shear strain. We discuss a model for ME coupling in the composites and explicit expressions for ME coefficients and resonance frequencies have been derived and estimated $\alpha_E$ values are found to be in general agreement with the data.

The multiferroic composites with piezoelectric quartz have the desired characteristics for use in magnetic sensors [9]. Ferroelectrics such as PZT or PMN-PT have crystallographic phase transitions and T$_c$ over the temperature range 20- 400$^0$C, and have ferroelectric and pyroelectric losses. Since trigonal $\alpha$-quartz is free of any phase transition up to 573$^0$C when it transforms to hexagonal $\beta$-quartz [10] and does not have ferroelectric hysteresis or pyroelectric losses encountered in PZT or PMN-PT, ME composites with quartz in particular are of interest for use as low-noise ultrasensitive magnetic sensors capable of operation at high temperatures. Details on ME studies and theory are provided in the sections that follow.

## 2. Experiment

Single crystal quartz substrates (obtained from MTI Corp., CA) were used. *X*–cut samples (*X* or direction-*1* being the piezoelectric axis) of dimensions 45x5x0.5 mm$^3$ were polished and Ti-Pt electrodes were deposited by rf sputtering. We first measured the parameters of importance, relative permittivity $\varepsilon_{11}$ and piezoelectric coupling coefficient $d_{11}$, for the X-cut samples and are shown in Fig.1. Data on $\varepsilon_{11}$ vs frequency *f* show a constant value of 6.8 up to 150 kHz and a series of peaks at higher frequencies corresponding to acoustic modes.

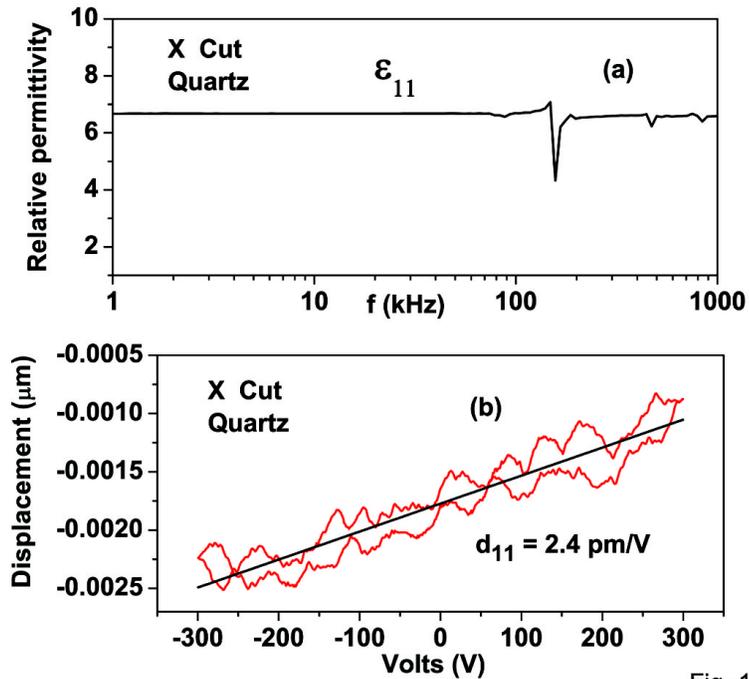

Fig.1: (a) Frequency dependence *f* of the relative permittivity $\varepsilon_{11}$ for X-cut quartz. (b) Piezoelectric displacement vs voltage *V* for X-cut quartz.

The piezoelectric displacement vs voltage *V* data in Fig.1 for the *X*-cut quartz is linear, as expected, and the estimated $d_{11}$ = 2.4 pm/V. The $\varepsilon_{11}$ and $d_{11}$ values are in agreement with reported values [13].

Permendur, a ferromagnetic alloy with 49% Fe, 49 % Co and 2% V, (obtained from the Institute of Technical Acoustics, Vitebsk, Belarus) and of dimensions 45 x 5 x 0.2 mm$^3$ was used in the composites with quartz. The permeability, a measure of magnetic field confinement, and magnetostriction $\lambda$ are important parameters for the ferromagnetic phase and were measured for permendur. The real part of the permeability $\mu'$ was measured over 10 Hz – 10 MHz on a toroid of and it decreased from 2300 at 10 Hz to 8 at 10 MHz. The magnetostriction $\lambda$ for permendur was measured with a strain gage. Figure 2 shows data on static field *H*-dependence of the in-plane $\lambda$ measured parallel to the field direction. With increasing *H*, one observes a rapid rise in $\lambda$ and it saturates at 70 ppm for *H* > 200 Oe. The in-plane $\lambda$ perpendicular to *H* (not shown here) was very small. The piezomagnetic coupling $q = d\lambda/dH$ estimated from the data in Fig.2 is also shown and $q$ increases with $H$ to a maximum of $1\times10^{-6}$/Oe at $H$ = 100 Oe. With further increase in $H$, $q$ shows a rapid drop and becomes zero for $H$ > 200 Oe.

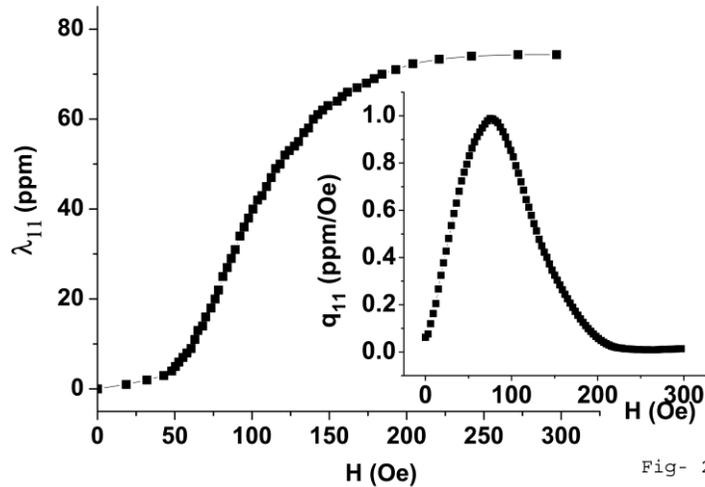

Fig.2: In-plane magnetostriction $\lambda_{11}$ measured parallel to the bias field *H* applied along the length of permendur. The piezomagnetic coefficient $q_{11} = d\lambda_{11}/dH$ is shown as a function of *H*.

Permendur of similar lateral dimensions as quartz and 0.2 mm in thickness were used in bilayers and trilayers. Platelets of P and quartz were bonded with a 2 μm thick epoxy layer that was cured at 40 C. For ME characterization the samples were subjected to a bias filed $H$ and ac field $\delta H$ = 1 Oe, both in-plane and parallel to the length of the sample, and the voltage $\delta V$ across the quartz layer (along direction *1*) was measured with a lock-in-amplifier. The ME voltage coefficient (MEVC) $\alpha_E = \delta V/(t\ \delta H)$, where $t$ is the thickness of quartz, was measured as a function of $H$ and frequency $f$ of the ac field.

3. Results

*3.1: ME effects in bilayers and trilayers of X-cut quartz and permendur*

We discuss first ME characterization of X-cut quartz and permendur bilayers. Figure 3 shows the static field $H$ and frequency $f$ dependence of the MEVC for quartz-P of lateral dimensions 45x5 mm$^2$ and a total thickness of 0.7 mm. The data on MEVC vs $H$ in Fig.3(a) are at room temperature and ac field $\delta H$ at $f$ = 20 Hz. A rapid increase in MEVC with $H$ to a maximum of 1.5 V/cm Oe at 30 Oe is observed and is followed by a decrease to a minimum for higher bias fields. The $H$-variation in MEVC in Fig.3(a) tracks the change in piezomagnetic coupling $q_{11}$ with $H$ for permendur shown in Fig.2. The observation of importance in Fig.3(a) is the MEVC values that are comparable to best values reported for ferroelectrics based composites as discussed later in Sec. 4.

The frequency dependence of the MEVC was measured with the bilayer rigidly clamped at one end and the results are shown in Fig.3(b) and (c). The bias field was set to the value corresponding to maximum MEVC in Fig.3(a). The data in Fig.3(b) shows a resonance in MEVC vs $f$ with a peak MEVC ≈ 185 V/cm Oe at 290 Hz due to bending oscillations in the bilayers.

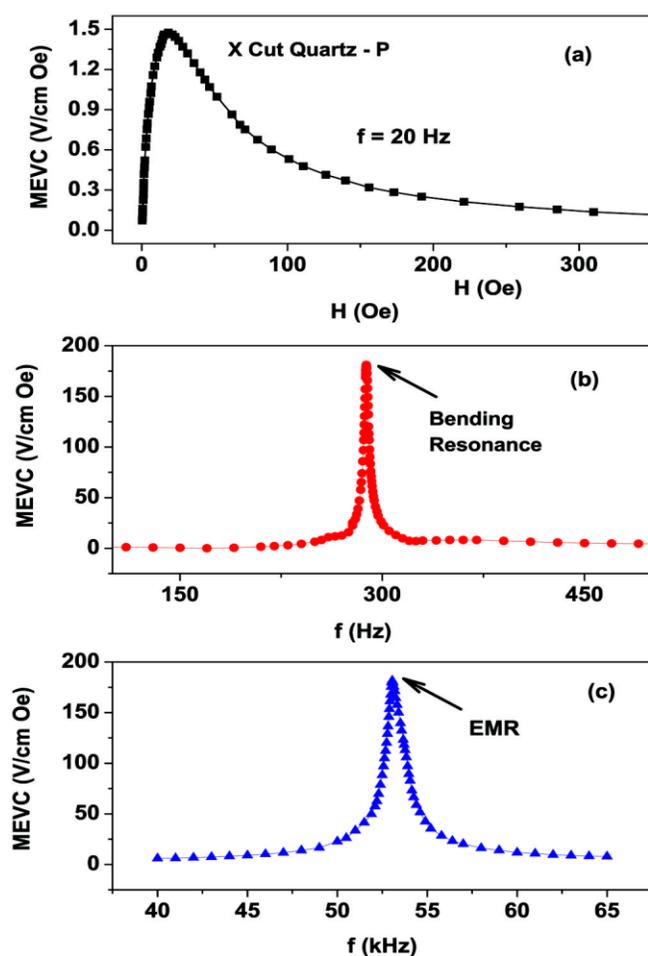

Fig.3 (a) Magnetoelectric voltage coefficient (MEVC) vs H for ac magnetic fields at 20 Hz for X-cut quartz-permendur bilayer. (b) Frequency f dependence of MEVC for the bilayer rigidly clamped at one end. The bias field H was set to the maximum MEVC in Fig.3(a). The resonance in MEVC occurs at bending mode for the bilayer. (c) Similar frequency dependence of MEVC showing a peak at electromechanical resonance (EMR) corresponding to longitudinal acoustic mode for the bilayer.

The *Q* value for the resonance is 145. Peaks at higher harmonics (not shown in Fig.3) of the bending mode were also observed, but with much lower MEVC than the fundamental mode. The resonance in Fig.3(c) occurs at 53 kHz and is due to the longitudinal acoustic mode for the bilayer. The peak-MEVC at EMR is 190 V/cm Oe, approximately of the same magnitude as for the bending mode.

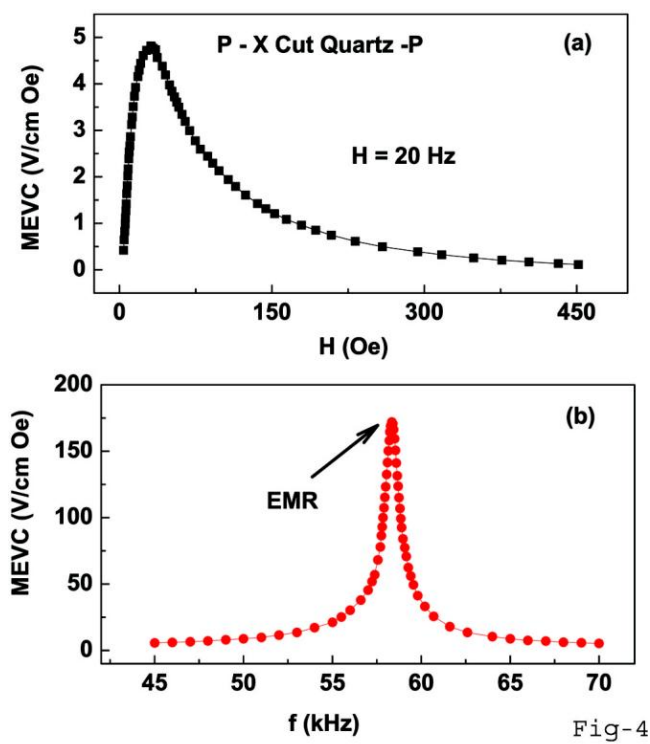

Fig.4: Similar (a) MEVC vs H and (b) MEVC vs f data for a trilayer of P-X-cut quartz-P. The resonance in Fig.3(b) occurs at electromechanical resonance corresponding to the longitudinal acoustic mode for the trilayer.

A symmetric trilayer structure of P-quartz-P was made by adding a 0.2 mm thick layer of permendur to the bilayer and similar ME characterization as for the bilayer were carried out. Figure 4 shows results of these measurements. The MEVC vs H data in Fig.4(a) shows a maximum MEVC of 5 V/cm Oe at 20 Hz and MEVC vs *f* data in Fig.4(b) shows a maximum MEVC of 175 V/cm Oe for electromechanical resonance (EMR) at 58 kHz. Thus the trilayer shows a factor of 3 enhancement in low-frequency MEVC compared to the bilayer. This can be attributed to the absence of bending strain in symmetric structures whereas such bending strains are present in bilayers due to sample asymmetry. Thus the highest off-resonance (low-frequency) MEVC is expected for the trilayer. Also the trilayers do not show any resonance enhancement at bending resonance since such modes are absent.

### *3.2: ME effects in trilayers of Y-cut quartz and permendur*

Next we consider ME coupling in a trilayer of permendur - Y-cut Quartz – Permendur. Quartz with the Y-axis along the thickness and of dimensions 45 x 5 x 0.5 mm$^3$ were used. The sample did not show strong ME response for in-plane longitudinal strain. This is illustrated in Fig.5(a) that shows the MEVC vs H data for bias field *H* and ac magnetic field $\delta H$ parallel to each other and along the X-direction. The MEVC is vanishingly small, with a peak value of 0.12 V/cm Oe, compared to ~ 4.8 V/cm Oe for the trilayers with X-cut quartz.

An electric field, however, can be induced along the Y-direction by face shear or thickness shear modes. As an example, we consider the face shear deformations. For exciting shear deformation in the XZ-plane, ac and dc magnetic fields should be applied to the sample in the Z- and X-direction, respectively. Data on MEVC vs H for this case is shown in Fig.5(b). A strong ME response under shear strain is evident from the data.

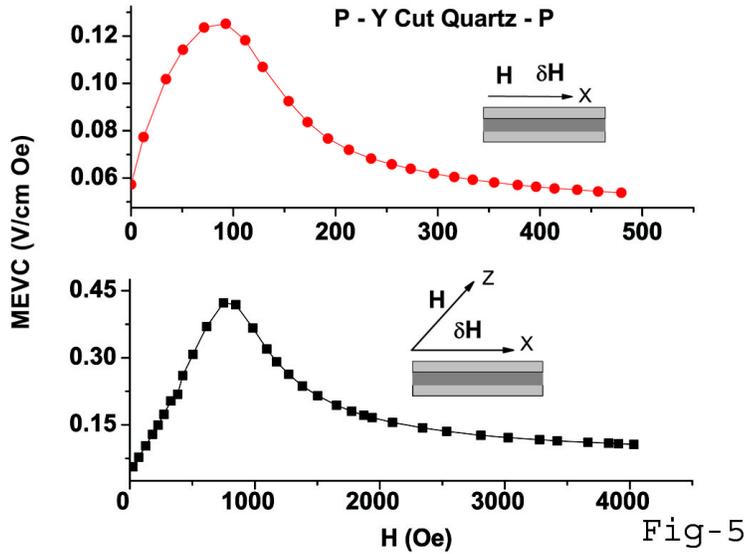

Fig-5

Fig. 5: Low-frequency MEVC vs H data a bilayer of Y-cut quartz and permendur for (a) bias field H and ac field δH parallel to X-direction and (b) H along Z-direction and δH along X-direction. Face-shear deformation leads to a strong ME coupling for this case.

### 4. Theory and Discussion

Next we discuss a theory for low-frequency and resonance ME effects in the composites with X-cut quartz and derive expressions for ME coefficients [14-16]. The sample is assumed to be a narrow thin plate with length $L$, width $w$ and thickness $t$ that obey the following condition: $t \ll w \ll L$. For X-cut quartz, the thickness direction is the piezoelectric axis $X$, (direction $1$) and the length is along $Y$ (direction $2$). The in-plane bias magnetic field and the ac magnetic are applied parallel to the direction $Y$. Under such conditions, the only nonzero stress components are the $Y$-components. The piezoelectric and piezomagnetic layers are assumed to be perfectly bonded together so that the piezoelectric layer restricts the deformation of the piezomagnetic layer in an external magnetic field. For symmetric trilayer structure, this stress component can be considered as constant in the low-frequency region over the sample volume. For asymmetric bilayer structures, the shear forces of each layer produce bending moments relative to the *Y*-axis since the forces are not applied centrally. As a result, a flexural deformation and a curvature appear. For taking into account the flexural deformations, the longitudinal axial strains of each layer should be considered as functions of the vertical coordinate and we assume the functions to be linear. Mechanical equilibrium conditions for total force and torque enable finding this stress. Substituting the value of stress into open electric circuit condition results in an analytical expression for ME voltage coefficients.

### *4.1. Low-frequency ME coupling*

We consider a trilayer of permendur-quartz-permendur. The ME coefficient $\alpha_E$ can be estimated by the following procedure. The total strain in the composite which is the sum of elastic and piezoelectric or elastic and magnetostrictive strains is estimated first. Then the strain-stress relationship is calculated for boundary conditions at the interface between phases. Finally the stress-electric field relationship is estimated for appropriate boundary conditions and the ME coefficient is estimated for open circuit conditions. An averaging method is used for deriving effective composite parameters. For the polarized piezoelectric phase with the symmetry ∞m, the following equations can be written for the strain and electric displacement:

$$^{p}S_i = {^{p}s_{ij}}\,{^{p}T_j} + {^{p}d_{ki}}\,{^{p}E_k}; \qquad (1)$$

$$^{p}D_k = {^{p}d_{ki}}\,{^{p}T_i} + {^{p}\varepsilon_{kn}}\,{^{p}E_n};$$

where $^{p}S_i$ and $^{p}T_j$ are strain and stress tensor components of the piezoelectric phase, $^{p}E_k$ and $^{p}D_k$ are the vector components of electric field and electric displacement, $^{p}s_{ij}$ and $^{p}d_{ki}$ are compliance and piezoelectric coefficients, and $^P\varepsilon_{kn}$ is the permittivity matrix. The magnetostrictive phase is assumed to have a cubic symmetry and is described by the equations:

$$^mS_i = {}^ms_{ij}\,{}^mT_j + {}^mq_{ki}\,{}^mH_k; \qquad (2)$$

$$^mB_k = {}^mq_{ki}\,{}^mT_i + {}^m\mu_{kn}\,{}^mH_n;$$

where $^mS_i$ and $^mT_j$ are strain and stress tensor components of the magnetostrictive phase, $^mH_k$ and $^mB_k$ are the vector components of magnetic field and magnetic induction, $^ms_{ij}$ and $^mq_{ki}$ are compliance and piezomagnetic coefficients, and $^m\mu_{kn}$ is the permeability matrix. Equation (2) may be considered in particular as a linearized equation describing the effect of magnetostriction. Assuming ideal in-plane mechanical connectivity between the two phases, the following expression for the ME voltage coefficients under 1-D approximation is obtained by solving Eqs.(1) and (2):

$$\frac{\delta E_1}{\delta H_2} = \frac{{}^mq_{11}\,{}^Pd_{11}/{}^P\varepsilon_{11}}{r\,{}^ms_{11} + {}^Ps_{11}}, \qquad (3)$$

where $\delta H_2$ is the applied ac magnetic field and $\delta E_1$ is the measured induced electric field, and $r = {}^Pt/{}^mt$ with $^Pt$ and $^mt$ denoting the thickness of piezoelectric and piezomagnetic layers, respectively. In Eq. 3, the electromechanical coupling factor is assumed to satisfy the condition that $^PK_{11}^2 = {}^Pd_{11}^2/{}^Ps_{11}\,{}^P\varepsilon_{11} \ll 1$. It should be noted that for quartz the piezoelectric coefficient $^Pd_{21} = -{}^Pd_{11}$.

For the permendur-quartz bilayer, the ME voltage coefficient should be calculated taking into account the flexural deformations. Our model in Ref. 15 enables deriving the explicit expression for ME voltage coefficient. For 1-D case and under the assumption $^PK_{11}^2 \ll 1$, we obtained

$$\frac{\delta E_1}{\delta H_2} = \frac{[1^{p}s_{11}+{}^{m}s_{11}r^{3}]^{m}q_{11}{}^{p}d_{11}/{}^{p}\varepsilon_{11}}{{}^{p}s_{11}[2r^{m}s_{11}(2+3r+2r^{2})+{}^{p}s_{11}]+{}^{m}s_{11}^{2}r^{4}}. \quad (4)$$

For theoretical estimates of MEVC for the bilayer and trilayer we used the following values of material parameters: ${}^{p}s_{11}$ =4.5×10$^{-12}$ m$^2$/N, ${}^{m}s_{11}$ =7.8×10$^{-12}$ m$^2$/N, ${}^{p}d_{11}$ =2.4×10$^{-12}$ m/V, and ${}^{p}\varepsilon_{11}/\varepsilon_0$=6.8 [11-13]. Data on $q_{11}$ vs $H$ in Fig.2 were used. Theoretical $H$ dependence of MEVC for the bilayer and trilayer are shown in Fig.6. For the trilayer, the estimated MEVC reaches a maximum value of 16 V/cm Oe and is approximately 4 times higher compared to MEVC for the bilayer. The low MEVC for the bilayer can be attributed to flexural deformations in the bilayer. Such deformations decrease the net axial stress component which determines the value of ME output. The measured MEVC vs $H$ are also shown in Fig.6 for comparison. A qualitative agreement between theory and data is evident. Theoretical and measured $H$-dependence of MEVC tracks each other. The measured peak-MEVC for the trilayers is 30% of the calculated values, whereas for the bilayer the measured value is 40% of the estimated value.

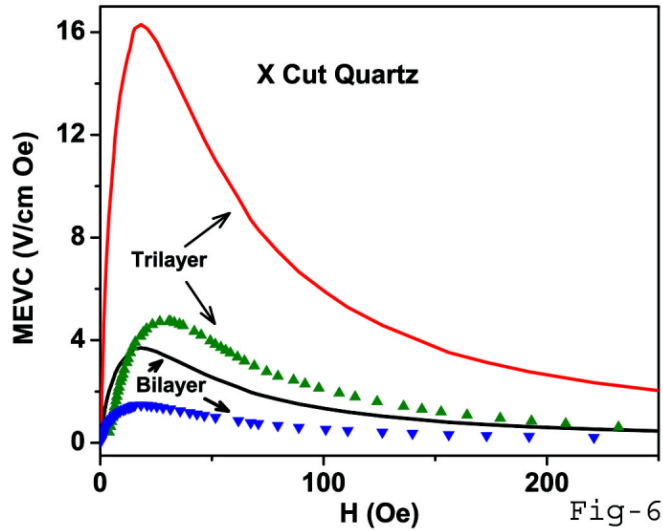

Fig. 6. Bias magnetic field dependence of low-frequency ME voltage coefficients for Permendur-X-cut Quartz bilayer and Permendur-X-cut Quartz-Permendur trilayer. Solid lines are theoretical values and the symbols are data.

Possible reasons for the discrepancy are (i) the assumption of 1-D model ($L \gg w \gg t$) for our samples and (ii) less-than ideal interface bonding. The bilayer has a total $t = 0.7$ mm and the thickness increases to 1.2 mm for the trilayer (versus lateral dimensions of 4.5 mm and 0.5 mm). Also the theory assumes ideal interface coupling. In our model in Ref.14, we introduced a parameter $k_b$ to include a less-than-ideal interface bonding. Although one could vary $k_b$ to achieve good agreement between theory and data in Fig.6, it is necessary to come up with an appropriate experimental procedure to measure the $k_b$ –value in order to have a physically meaningful comparison between theory and data.

### *4.2 ME coupling at bending mode*

Next we consider ME coupling under small-amplitude flexural oscillations of a bilayer. In our previous modelling work in Ref.15, we showed that a bilayer rigidly clamped at one end has the lowest mode frequency for longitudinal bending oscillations. In the present case, the bending deflection should obey the equation of motion as in Eq.1 in Ref.15. Assuming a periodic harmonic oscillation associated with the bending mode (Eq.5 in Ref.15), we used the boundary conditions (Eq.17 in Ref.15) that the deflection and its derivative vanish at clamped end and that the rotational moment and transverse force must vanish at free end. Under our assumptions $^p K_{11}^2 \ll 1$ and $^m K_{11}^2 \ll 1$ [$^m K_{11}^2 = {^m q_{11}^2}/{(^m s_{11} {^m \mu_{11}})}$], the resonance condition for 1-D case is $cosh(kL) \cdot cos(kL) = -1$ where $k$ is wave number. The fundamental resonance frequency can then be expressed as

$$f_r = \frac{1.758}{\pi L^2} \sqrt{\frac{D}{^p\rho\, ^p t + {^m\rho}\, ^m t}}. \tag{5}$$

Here $D$ is the cylindrical stiffness and is defined as $D = \frac{[z_0^3 - (z_0 - {}^p t)^3]}{3 {}^p s_{11}} + \frac{[(z_0 + {}^m t)^3 - z_0^3]}{3 {}^m s_{11}}$ and $z_0$ is the distance of a plane along the sample thickness (direction *1*) where the axial force vanishes and can be estimated from (Eq.2 in Ref15) $z_0 = \frac{1}{2} \frac{{}^p Y \, {}^p t^2 - {}^m Y \, {}^m t^2}{{}^p Y \, {}^p t + {}^m Y \, {}^m t}$. Here $Y$ is the Young's modulus and $\rho$ is the density. The peak ME voltage coefficient at bending mode frequency is found to be (from Eq.10 in Ref.15)

$$\frac{\delta E_1}{\delta H_2} = \frac{0.0766 Q_b}{D {}^p s_{11} {}^m s_{11}} {}^m t ({}^m t + 2 z_0)(2 z_0 - {}^p t) {}^m q_{11} \, {}^p d_{11} / {}^p \varepsilon_{11} \tag{6}$$

where $Q_b$ is the quality factor for bending resonance.

Using the material parameters for quartz and permendur [11,13] and the calculated $Q_b = 144$ from data in Fig.3, we estimated the bending resonance frequency $f_r = 265$ Hz and peak MEVC = 202 V/cm Oe. These values are in very good agreement with measured $f_r = 290$ Hz and MEVC = 185 V/cm Oe (Fig.3).

### *4.3 ME coupling at axial mode of electromechanical resonance*

Finally we consider small-amplitude longitudinal acoustic oscillations of the layered structures formed by the magnetostrictive and piezoelectric phases. Such oscillations, present in both bilayers and trilayers, lead to enhancement at electromechanical resonance at frequencies much higher than for bending modes in bilayers. Here we apply our previous model in Ref.16 for a bilayer in the shape of a circular disk. The displacement should obey the equations of media motion provided in Ref.16. We modified the strain tensor and relevant expressions (Eqs.1-6 in Ref.16) for the present case of 1-D bilayer that is free at both ends [16]. Under the assumption $^PK_{11}^2 \ll 1$, the fundamental EMR frequency (by setting $\Delta_a = 0$ in Eq.6 of Ref.16) is given by

$$f = \frac{1}{2L}\sqrt{\frac{^Ps_{11} + r\,^ms_{11}}{^Ps_{11}\,^ms_{11}(r\,^P\rho + ^m\rho)}}, \tag{7}$$

and the peak ME voltage coefficient at axial mode frequency (from Eq.7 in Ref.16) is

$$\frac{\delta E_1}{\delta H_2} = \frac{8Q_a}{\pi^2}\frac{^mq_{11}\,^Pd_{11}/^P\varepsilon_{11}}{r\,^ms_{11} + ^Ps_{11}}, \tag{8}$$

where $Q_a$ is the quality factor for the EMR resonance.

For the bilayer of quartz and permendur, using appropriate material parameter values and $Q_a=66$ (from data in Fig.3c) the estimated EMR frequency $f_r$=51.5 kHz and MEVC at resonance is 206 V/cm Oe. These values are in good agreement with the measured values of $f_r$ = 53 kHz and MEVC = 190 V/cm Oe in Fig.3(c).

It should be noted that Eqs. (7) and (8) for resonance frequency and ME voltage coefficient are valid for both bilayer and trilayer structures. For the trilayer of permendur-quartz-permendur, the following values of resonance frequency and peak ME voltage coefficient at EMR frequency are obtained: $f_r$ = 48.6 kHz and MEVC=291 V/cm Oe. The calculated $f_r$ is 20% smaller than the measured value of 58 kHz and the estimated MEVC is much higher than the measured value of 175 V/cm Oe.

### *4.4: Discussion*

Next we discuss the low-frequency and resonance MEVC for X-cut quartz-permendur bilayers and trilayers in Figs. 3 and 4. The ME voltage coefficient for in-plane magnetic fields is proportional to $d_{11}/(\varepsilon_0 \varepsilon_{11})$ for quartz and LGT and $d_{13}/(\varepsilon_0 \varepsilon_{33})$ for PZT or PMN-PT. Using the appropriate values for the relative permittivity $\varepsilon_r = \varepsilon_{11}$ and $\varepsilon_{33}$ and the piezoelectric coupling coefficients [11-13], we estimated the ratio $d/\varepsilon_r$. Figure 7 shows the low-frequency and resonance ME coefficients for permendur based composites as a function of $d/\varepsilon_r$. It is well known that MEVC is critically dependent on the composite geometry and dimensions (length, width and thickness of ferromagnetic and ferroelectric phases) [17,18]. The MEVC-values used here for comparison were measured on samples with approximately same dimensions [9].

Consider first the low-frequency ME effects. It is clear from the data in Fig.6 that the MEVC at 20 Hz shows a general increase with increasing $d/\varepsilon_r$ for both bilayers and trilayers with the maximum value measured for piezoelectric-based composites. Although $d/\varepsilon_r$ is higher for quartz than for LGT, the MEVC does not show the expected maximum ME response for quartz-P. The highest MEVC at 20 Hz is measured for trilayers and is a factor of 3-10 higher than for bilayers. This can be attributed to bending strain in bilayers in a magnetic field due to assymmetry. Such a bending strain is absent in symmetric trilayers and the highest MEVC is measured for the trilayer as expected. The low-frequency MEVC values in Fig.6 for quartz-P bilayers and trilayers are orders of magnitude higher than reported values for bulk ferrite-piezoelectric composites, for bilayers and trilayers of ferrite-PZT and lanthanum manganite-PZT [1-3]. The maximum MEVC at 20 Hz in Fig.6 compares favorably with MEVC of 3 - 52 V/cm Oe at 1 kHz for Metglas composites with PZT fibers and single crystal PMN-PT or PZN-PT [19,20].

Data on resonance MEVC in Fig.7 are considered next. The ME coefficient for EMR in trilayers increases with increasing $d/\varepsilon_r$ and is maximum for P-quartz-P. The resonance-MEVC under EMR in Fig.7 is one to two orders of magnitude higher than the low-frequency MEVC.

The highest MEVC ~ 1100 V/cm Oe under EMR is reported for Metglas-PMN-PT [21]. For bending resonance in bilayers, the data in Fig.6 shows a general increase in MEVC with $d/\varepsilon_r$ but the maximum MEVC is measured for LGT-P. The highest MEVC of 737 V/cm Oe under bending resonance at 753 Hz is reported for AlN/Fe-Co-Si-B [7].

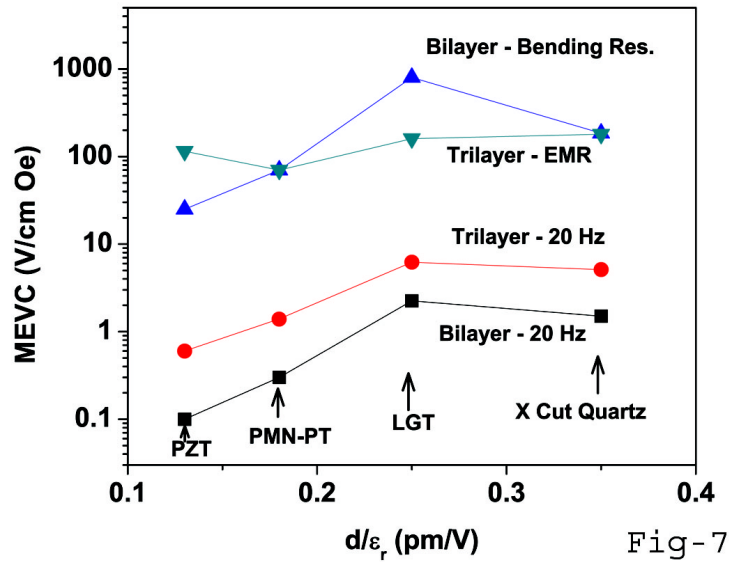

Fig.7: Low-frequency and resonance MEVC as a function of the ratio of the piezoelectric coupling coefficient $d$ to the relative permittivity $\varepsilon_r$ for bilayers and trilayers of permendur and ferroelectric PZT or PMN-PT and piezoelectric LGT or quartz.

5. **Conclusion**

The mechanical strain mediated magneto-electric coupling is studied in layered composites of piezoelectric quartz and magnetostrictive permendur. Bilayers and trilayers of single crystal X-cut quartz and permendur are characterized in terms of ME coupling at low-frequency and bending resonance and electromechanical resonance. It is shown that the MEVC for X-cut quartz-permendur is higher than for similar composites with ferroelectrics such as PZT or PMN-PT and is attributed to higher value of the ratio $d/\varepsilon$ (piezoelectric coupling coefficient $d$, permittivity $\varepsilon$). A model is developed for the ME coupling and the estimated MEVC are in general agreement with the data.

**Acknowledgments**

The research at Oakland University was supported by a grant from the National Science Foundation (DMR-0902701).